\documentclass{ifacconf}

\usepackage{graphicx}
\usepackage[dvips]{epsfig}

\usepackage{amssymb,amsmath,amsfonts,subfigure,enumerate}
\usepackage{multirow}
\usepackage{bbm}
\usepackage{natbib}

\def\la{\lambda}
\def\La{\Lambda}

\def\ve{\varepsilon}

\def\g{\mathcal G}

\def\r{\mathbb R}

\def\be{\beta}

\def\S{\mathfrak S}

\def\diag{\mathop{\rm diag}\nolimits}

\newcommand{\dfb}{\stackrel{\Delta}{=}}
\def\ones{\mathbf{1}}

\def\fj{\mathop{\mathbf{FJ}}\nolimits}
\def\cfj{\mathop{\mathbf{CFJ}}\nolimits}

\def\be{\begin{equation}}
\def\ee{\end{equation}}
\def\ben{\begin{equation*}}
\def\een{\end{equation*}}

\begin{document}

\begin{frontmatter}
\title{Opinion evolution in time-varying social influence networks with prejudiced agents}

\thanks[footnoteinfo]{
Partial funding was provided by NWO (vidi-438730), ERC (grant ERC-StG-307207),
CNR International Joint Lab COOPS, Russian Federation President's Grant MD-6325.2016.8 and 
RFBR, grants 17-08-01728, 17-08-00715 and 17-08-01266. Theorem~2 is obtained under sole support 
of Russian Science Foundation grant 14-29-00142. E-mails: {\tt\small anton.p.1982@ieee.org, m.cao@rug.nl, friedkin@soc.ucsb.edu}
}

\def\theaddress{\arabic{address}}

\author[DCSC,IPME,ITMO]{Anton V. Proskurnikov}
\author[CNR]{Roberto Tempo}
\author[RUG]{Ming Cao}
\author[UCSB]{Noah E. Friedkin}

\address[DCSC]{Delft Center for Systems and Control (DCSC), Delft University of Technology, The Netherlands}
\address[IPME]{Institute for Problems of Mechanical Engineering (IPME RAS), St. Petersburg, Russia}
\address[ITMO]{ITMO University, St. Petersburg, Russia}
\address[CNR]{CNR-IEIIT, Politecnico di Torino, Torino, Italy}
\address[RUG]{Engineering and Technology Institute (ENTEG), University of Groningen, The Netherlands}
\address[UCSB]{Center for Control, Dynamical Systems and Computation, University of California Santa Barbara, Santa Barbara, USA}

\begin{abstract}
Investigation of social influence dynamics requires mathematical models that are
``simple'' enough to admit rigorous analysis, and yet sufficiently ``rich'' to capture salient features of social groups. Thus, the mechanism of iterative opinion pooling from~\citep{DeGroot}, which can explain the generation of consensus, was elaborated in~\citep{FriedkinJohnsen:1999} to take into account individuals' ongoing attachments to their initial opinions, or \emph{prejudices}.
The ``anchorage'' of individuals to their prejudices may disable reaching consensus and cause disagreement in a social influence network. Further elaboration of this model may be achieved by relaxing its restrictive assumption of a time-invariant influence network. During opinion dynamics on an issue, arcs of interpersonal influence may be added or subtracted from the network, and the influence weights assigned by an individual to his/her neighbors may alter. In this paper, we establish new important properties of the \citep{FriedkinJohnsen:1999} opinion formation model, and also examine its extension to time-varying social influence networks.
\end{abstract}

\end{frontmatter}


\section{Introduction}

During the past decades, there has been a substantial growth of interest in dynamics of social influence networks and opinion formation mechanisms in them. In contrast to the recent research emphasis on multi-agent consensus and coordination, models are being advanced that explain observed behaviors of social groups such as disagreement, polarization, and conflict~\citep{Friedkin:2015,ProTempo2017-1}. An explanatory network science is advancing on the structural properties of social networks~\citep{WassermanFaustBook,EasleyKleinberg}
and some special dynamical processes over these networks, e.g. epidemic spread~\citep{Newman:2003}. At the same time, there is a growing recognition that systems and control theories may substantially broaden the scope of our understanding of the definitional problem of sociology---the coordination and control of social systems \citep{Friedkin:2015}.

System-theoretic examination of social dynamics requires mathematical models that are capable of capturing the complex behavior of a social group yet simple enough to be rigorously examined. In this paper, we deal with one such model, proposed by Friedkin and Johnsen~\citep{FriedkinJohnsen:1999,FriedkinJohnsenBook,Friedkin:2015} and henceforth referred to as the FJ model. The FJ model extends the idea of iterative ``opinion pooling''~\citep{DeGroot} by assuming that some agents are \emph{prejudiced}. These agents have some level of ``anchorage'' on their initial opinions (\emph{prejudices}) and factor them into any iteration of their opinions. Similar to continuous-time clustering protocols with ``informed'' leaders~\citep{XiaCao:11}, the heterogeneity of the prejudices and its linkage to individuals' susceptibilities to interpersonal influence may lead to persistent disagreement of opinions and outcomes such as
polarization and clustering. With the FJ model, the clustering of opinions does not require the existence of repulsive couplings, or 
``negative ties'' among individuals~\citep{Flache:2011,Altafini:2013,ProMatvCao:2016,XiaCaoJohansson:16} whose ubiquity in interpersonal 
interactions is still waiting for supporting experimental evidence~\citep{TakacsFlacheMas:16}.
Unlike models with discrete opinions~\citep{Castellano:2009} and bounded confidence models~\citep{Krause:2002,WeisbuchDeffuant:2005,Blondel:2009},
 the FJ model describes the opinion evolution by \emph{linear} discrete-time equations, and is thus much simpler for mathematical analysis. 
At the same time, the FJ model has been confirmed by experiments with real social groups~\citep{FriedkinJohnsenBook,FriedkinJiaBullo:2016}. The FJ model is closely related to the PageRank algorithm~\citep{FriedkinJohnsen:2014,ProTempoCao16-1} and has been given some
elegant game-theoretic and electric interpretations~\citep{Bindel:2011,GhaderiSrikant:2014,FrascaIshiiTempo:2015}. In the recent works~\citep{Parsegov2017TAC,ProTempo2017-1} necessary and sufficient conditions for the stability of the FJ model has been established; these conditions also provide convergence ``on average'' of its decentralized gossip-based counterpart~\citep{FrascaTempo:2013,FrascaTempo:2015,FrascaIshiiTempo:2015}.
A multidimensional extension of the FJ model has been used to describe the evolution of  \emph{belief systems}~\citep{Parsegov2017TAC,FriedkinPro2016}, representing invidiuals' positions on several \emph{mutually dependent} issues.

In this paper, we further develop the mathematical theory of the FJ model, obtaining explicit estimates for its convergence speed.
We also examine an extension of the classical FJ model, describing a natural \emph{time-varying} social influence process. Such an extension is important since during opinion dynamics on
an issue, arcs of interpersonal influence may be added or subtracted from the network, and the influence weights
assigned by an individual to his/her neighbors may alter. An example of such an evolution is the dynamics of individuals' \emph{reflected appraisals}~\citep{Bullo:2013,FriedkinJiaBullo:2016,ChenLiuXuBasar:2016}.

\section{Preliminaries and notation}
We denote matrices with capital letters $A=(a_{ij})$, using lower case letters for their scalar entries and vectors.
The symbol $\ones_n$ denotes the column vector of ones $(1,1,\ldots,1)^{\top}\in\r^n$, and $I_n$ is the identity $n\times n$ matrix. For two vectors $x,y\in\r^n$ we write $x\le y$ if $x_i\le y_i\forall i$. The spectral radius of a square matrix $A$ is denoted by $\rho(A)$, the matrix is \emph{Schur} stable if $\rho(A)<1$.
A non-negative matrix $A$ is \emph{substochastic} if $\sum_ja_{ij}\le 1$ for any $i$. Any such matrix has $\rho(A)\le 1$ due to the Gershgorin disk theorem~\citep{HornJohnsonBook1985}. A substochastic matrix $A$ is \emph{stochastic} if $\sum_ja_{ij}=1\,\forall i$; when $A$ is sized $n\times n$, the stochasticity implies that $A\ones_n=\ones_n$ and $\rho(A)=1$.

A (weighted directed) \emph{graph} is a triple $\mathcal G=(\mathcal V,\mathcal E, W)$, where $\mathcal V=\{v_1,\ldots,v_n\}$ stands for the set of \emph{nodes},
$\mathcal E\subseteq \mathcal V\times \mathcal V$ is the set of \emph{arcs}, and $W$ is a (weighted) $n\times n$ \emph{adjacency matrix},
i.e. $w_{ij}>0$ when $(i,j)\in\mathcal E$ and otherwise $w_{ij}=0$. Henceforth we assume that $\mathcal V=\{1,2,\ldots,n\}$ and thus the graph $\g=\g(W)$ is uniquely defined by its adjacency matrix $W$. We denote an arc $(i,j)\in\mathcal E$ by $i\mapsto j$ and call the value $w_{ij}$ its \emph{weight}.
A chain of arcs $i_0\mapsto i_1\mapsto\ldots \mapsto i_{r-1}\mapsto i_r$ is a \emph{walk} of length $r$
from node $i_0$ to node $i_r$.

\section{The Friedkin-Johnsen model}

The FJ model describes a network of \emph{social influence}~\citep{FriedkinJohnsenBook}, consisting of $n$ individuals, or social \emph{agents} indexed 1 through $n$. The agents opinions are represented by scalars $x_i\in\r$, constituting the vector of opinions $x=(x_1,\ldots,x_n)^{\top}$. The process of social influence is described by two matrices: a stochastic matrix of \emph{interpersonal influences} $W\in \r^{n \times n}$ and a diagonal matrix $\La=\diag(\la_{11},\ldots,\la_{nn})$ of individual \emph{susceptibilities} $\la_{ii}\in [0;1]$ to the interpersonal influence.
At each step, the vector of opinions changes as follows
\begin{equation}\label{eq.fjmodel}
x(k+1) = \Lambda W x(k)+(I_n-\Lambda)u,\quad k=0,1,\ldots.
\end{equation}
The elements $u_i$ of the constant vector $u$ stand for the agents' \emph{prejudices}; the original FJ model~\citep{FriedkinJohnsen:1999,Friedkin:2015} assumed that $u_i=x_i(0)$.

In the special case where $\La=I_n$ the model~\eqref{eq.fjmodel} reduces to DeGroot's iterative ``opinion pooling''~\citep{DeGroot}, providing a discrete-time \emph{consensus algorithm}~\citep{RenBeardBook}. At each step, an agent sets its new opinion to be the convex combination of its own and others' opinions
\be\label{eq.degroot-scalar}
x_i(k+1)=\sum_{j=1}^nw_{ij}x_j(k)\,\forall i\Longleftrightarrow x(k+1)=Wx(k).
\ee
The influence weight $w_{ij}$ shows the contribution of $j$th opinion on each stage to the $i$th opinion on the next stage.

The FJ model~\eqref{eq.fjmodel} also employs the mechanism of convex combination, allowing some agents to be prejudiced. If $\la_{ii}<1$ then agent $i$ is ``attached'' to its prejudice $u_i$ and factors it into any opinion iteration, replacing~\eqref{eq.degroot-scalar} by
\be\label{eq.fj-scalar}
x_i(k+1)=\la_{ii}\sum_{j=1}^nw_{ij}x_j(k)+(1-\la_{ii})u_i\,\forall i.
\ee
When $\la_{ii}=1$, the $i$th agent's opinion is formed by the DeGroot mechanism~\eqref{eq.degroot-scalar}, otherwise its prejudice influences each stage of the opinion iteration.
Agent $i$ with $\la_{ii}=0$ is ``totally prejudiced'' and its opinion is static $x_i(k)\equiv u_i$.

Under the assumption $u_i=x_i(0)$, adopted in the FJ model, any agent with $w_{ii}=1$ (and thus $w_{ij}=0\,\forall j\ne i$) retains its opinion constant $x_i(k)=u_i$ independent of $\la_{ii}$, and one may suppose, without loss of generality, that
\be\label{eq.non-degen}
w_{ii}=1\Longleftrightarrow \la_{ii}=0.
\ee
In the original model from~\citep{FriedkinJohnsen:1999} an even stronger \emph{coupling condition}  $w_{ii}=1-\la_{ii}\,\forall i$ was adopted for parsimony in the model's empirical applications. In this paper, we do not assume this condition to hold, so $\La$ and $W$ are independent except for the non-degeneracy condition~\eqref{eq.non-degen}. Notice that each FJ model corresponds to the substochastic matrix $A=\La W$; for the models satisfying~\eqref{eq.non-degen} this
correspondence is one-to-one. A substochastic matrix $A$ is decomposed as $A=\La W$, where 
\ben
\la_{ii}=\sum_{j}a_{ij}\quad\text{and}\quad
w_{ij}=\begin{cases}
a_{ij}/\la_{ii},&\la_{ii}>0,\\
1, &i=j\,\text{and}\,\la_{ii}=0,\\
0, &i\ne j\,\text{and}\,\la_{ii}=0.
\end{cases}
\een
The stability criteria for FJ models may thus be reformulated for substochastic matrices, and vice versa.

For us it will be convenient to discard the standard assumption $x(0)=u$ and consider $u$ as some constant external ``input'', independent of the initial opinion\footnote{Individuals prejudices may be explained~\citep{FriedkinJohnsen:1999} by the system ``history'', e.g. the effect of some exogenous factors, which influenced the community in the past. This motivates to introduce the explicit relation between the prejudice and initial condition of the
social system $u=x(0)$. However, the prejudices can also be some non-trivial functions of the initial conditions $u=u(x(0))$ or
be caused by external factors that are not related to the system's history, e.g. some information spread in social media.} $x(0)$.

A central question concerned with the FJ dynamics~\eqref{eq.fjmodel} is its \emph{convergence} of opinion vectors to
a finite limit
\be\label{eq.x-inf0}
x^{\infty}=\lim_{k\to\infty}x(k).
\ee
A sufficient condition for convergence is the \emph{Schur stability}: if $\rho(\La W)<1$ then the opinions converge to
\be\label{eq.x-inf}
x^{\infty}=Vu,\quad V=(I-\La W)^{-1}(I-\La).
\ee
It is known~\citep{Friedkin:2015} that for any Schur stable FJ model the matrix $V$ is stochastic and, obviously, $x^{\infty}$ from~\eqref{eq.x-inf} is the only equilibrium of the system~\eqref{eq.fjmodel}.
Generally, the Schur stability is not necessary for convergence, e.g. the DeGroot model~\eqref{eq.degroot-scalar} is never stable but converges when e.g. $W$ is \emph{primitive} (i.e. irreducible and aperiodic)~\citep{DeGroot,GantmacherVol2}.

Henceforth we are primarily interested in Schur stable FJ models, where the steady opinion is unique and given by~\eqref{eq.x-inf}. The Schur stability is a ``generic'' condition if at least one prejudiced agent exists, and holds, for instance, for a strongly connected influence networks~\footnote{This property can be also reformulated as follows: an irreducible substochastic matrix is Schur stable~\cite[Exercise 8.3.7]{Meyer2000Book}.}, as implied by the following lemma~\citep{Parsegov2017TAC}.
\begin{lem}\label{lem.stab0}
The matrix $\La W$ is Schur stable if and only if each node in the graph $\g(W)$ either belongs to the set
\be\label{eq.s}
S_{\La}=\{i:\la_{ii}<1\}
\ee
or connected to a node from $S_{\La}$ by a walk, i.e. any agent is either prejudiced or influenced by a prejudiced individual.
\end{lem}

\section{Schur stable FJ models: opinion clustering and convergence speed}\label{sec:stationary}

In this section, we derive some advanced properties of Schur stable FJ models~\eqref{eq.fjmodel},
satisfying the condition from Lemma~\ref{lem.stab0}.
We answer the following two questions, related to such models' dynamics:
\begin{itemize}
\item do the final opinions $x^{\infty}$ reach consensus or disagree?
\item what is the convergence speed in~\eqref{eq.x-inf0}?
\end{itemize}

\subsection{Consensus and disagreement in the FJ model}

One can expect that for a general FJ model the consensus of the steady opinions $x^{\infty}_1,\ldots,x^{\infty}_n$ typically should disagree, whereas their consensus is an exceptional situation.
This is confirmed by the following consensus criterion.
\begin{thm}\label{thm.cons}
Let the FJ model~\eqref{eq.fjmodel} be stable. Then the consensus of final opinions $x^{\infty}_1=\ldots=x^{\infty}_n$ is reached if and only if $u_i=u_0$ for some $u_0\in\r$ and any $i\in S_{\La}$. In this case $x^{\infty}_i=u_0\,\forall i$. This holds e.g. when $S_{\La}=\{i\}$ has only one element, i.e. only one agent is prejudiced.
\end{thm}

The values $u_i$, where $i\not\in S_{\La}$ and thus $1-\la_{ii}=0$, obviously do not influence the value of $x^{\infty}$ and may be arbitrary. Note that when consensus is not established, the number of ``clusters'' in the vectors $u$ and $x^{\infty}$ do not correlate. For instance, if $u_1=1$ and $u_i=0\,\forall i>1$ then $x^{\infty}$ is the first row of $V$ and its elements are usually all different.

\subsection{Convergence speed of the FJ model}

In this subsection we give an explicit estimate of the spectral radius $\rho=\rho(\La W)$, which also determines the convergence speed in~\eqref{eq.x-inf0}:
$|x(k)-x_*|=O(\rho^k)$ as $k\to\infty$.
We start with introducing some definitions and notation.
\begin{defn}
An arc $i\mapsto j$ in $\g(W)$ with the weight $w_{ij}\ge\ve$ is said to be an $\ve$-arc. An $\ve$-walk in the graph is a walk constituted by $\ve$-arcs.
Given a set $S\subseteq\{1,\ldots,n\}$ and node $i$, let $d_{\ve}(i,S,W)$ stand for the length of the \emph{shortest} $\ve$-walk from $i$ to $S$.
By definition, $d_{\ve}(i,S,W)=0$ for any $i\in S$ and $d_{\ve}(i,S,W)=\infty$ if no $\ve$-walk from $i$ to $S$ exists.
\end{defn}

For any diagonal matrix $\La=\diag(\la_{11},\ldots,\la_{nn})$ with $0\le\La\le I_n$ and $\delta>0$, we introduce the set of indices
\be\label{eq.sdelta}
S_{\La}^{\delta}=\{i:\la_{ii}\le 1-\delta\}\subseteq S_{\La}.
\ee

\begin{defn}
The FJ model (or the pair $(\La,W)$) belongs to the class $\fj[\delta,\ve,s]$ if $d_{\ve}(i,S_{\La}^{\delta},W)\le s$ for any node $i=1,\ldots,n$.
Here $\delta,\ve>0$ are real and $s\ge 0$ is an integer.
\end{defn}

Any FJ model, belonging to $\fj[\delta,\ve,s]$ with $\delta,\ve>0$, is Schur stable due to Lemma~\ref{lem.stab0}. On the other hand, any Schur stable FJ model belongs to
$\fj[\delta_0,\ve_0,n-1]$, where
\be\label{eps-0}
\ve_0\dfb\min\{w_{ij}:w_{ij}>0\},\quad \delta_0\dfb 1-\max\limits_{i\in S_{\La}}\la_{ii},
\ee
since $S_{\La}^{\delta_0}=S_{\La}$ and any walk in $\g(W)$ is an $\ve_0$-walk.

The following theorem gives an explicit estimate for the spectral radius $\rho(\La W)$ of a Schur stable FJ model~\eqref{eq.fjmodel}.
\begin{thm}\label{thm.stab}
For any FJ model~\eqref{eq.fjmodel} from the class $\fj[\delta,\ve,s]$ one has $\rho(\La W)\le \rho_*(\delta,\ve,s)\dfb\left[1-\delta\ve^s\right]^{1/(1+s)}$.
\end{thm}

For the case of undirected graph $\g(W)$ and special influence weights of arcs a similar estimate for the convergence speed has been obtained in~\citep{GhaderiSrikant:2014}. Unlike this paper, Theorem~\ref{thm.stab} deals with a general FJ model, where the matrix $W$ can be arbitrary.
\begin{cor}\label{cor.stab}
For a stable FJ model~\eqref{eq.fjmodel}, let $\ve_0,\delta_0$ be defined by~\eqref{eps-0}. Then
$\rho(\La W)\le \left[1-\delta_0\ve_0^{n-1}\right]^{1/n}$.
\end{cor}

Although the estimate from Theorem~\ref{thm.stab} is just an upper bound for $\rho(\La W)$,
this bound proves to be tight for special types of graphs.
For instance, if $\La=(1-\delta)I_n$ then $S_{\La}^{\delta}=\{1,\ldots,n\}$, $s=0$ and hence
$\rho(\La W)=(1-\delta)$ for any $W$. Another example is the cycle graph $1\mapsto 2\mapsto 3\mapsto\ldots\mapsto n\mapsto 1$, where the weights of arcs are equal to $\ve=1$.
If $\la_{11}=1-\delta$ and $\la_{22}=\ldots=\la_{nn}=1$ then $s=n-1$ and
$$
\La W=
\left[\begin{array}{ccc|c}
0 & \cdots & 0 & 1-\delta\\\hline
\multicolumn{3}{c|}{\multirow{3}{*}{\raisebox{-10pt}{$I_{n-1}$}}} & 0\\
& & & \raisebox{5pt}{\vdots} \\
& & & 0
\end{array}\right],\, \rho(\La W)=(1-\delta)^{1/n}.
$$

\section{Time-varying FJ model}\label{sec:time-var}

A principal limitation of the standard FJ model~\eqref{eq.fjmodel} is the \emph{time invariance} of social influence: the matrices $\La$ and $W$ remain constant. In real social groups
the structures of social influence may evolve over time as the interpersonal ties may emerge and disappear; even if their graph remains constant, the influence weights $w_{ij}$ and susceptibilities $\la_{ii}$ may change. One of the models, describing the evolution of the matrix $W$, is the dynamics of \emph{reflected appraisals}~\citep{Bullo:2013,ChenLiuXuBasar:2016,FriedkinJiaBullo:2016}, where the self-confidence of a person depends on how he/she is evaluated by the others. In this section we consider a time-varying extension of the FJ model and study its properties.

The time-varying FJ model (TVFJ) is as follows
\be\label{eq.fjmodel-tv}
x(k+1)=\La(k)W(k)x(k)+[I_n-\La(k)]u.
\ee
We assume that the matrices $\La(k),W(k)$ on each stage of the opinion evolution are known and have the same structure, as for the classical model~\eqref{eq.fjmodel}, i.e. $\La(k)$ is diagonal, $0\le \La(k)\le I_n$ and $W(k)$ is stochastic. Given the initial condition $x(0)=x^0$ and the prejudice vector $u$, let $x(k|x^0,u)$ stand for the solution of~\eqref{eq.fjmodel-tv}.
The averaging mechanism of~\eqref{eq.fjmodel-tv} provides several useful properties.
\begin{lem}\label{lem.simple} Any model~\eqref{eq.fjmodel-tv} has the following properties:
\begin{enumerate}
\item if $x^0=u=u_*\ones_n$, then $x(k|x^0,u)=u_*\ones_n\,\forall k$;
\item if $x^0,u\in [m,M]^n$, then $x(k|x^0,u)\in [m,M]^n\,\forall k$;
\item more generally, if $x^0\le x^1$ and $u\le u^1$, then $x(k|x^0,u)\le x(k|x^1,u^1)\,\forall k$;
\item for any ``perturbations'' $\delta x^0,\delta u\in [m,M]^n$ one has $x(k|x^0+\delta x^0,u+\delta u)-x(k|x^0,u)\in [m,M]^n$.
\end{enumerate}
Here $m,M,u_*$ stand for some real scalars.
\end{lem}

Applied for $M=-m=\ve$, statement 4) in Lemma~\ref{lem.simple} implies \emph{robustness} of the trajectories against small perturbations in $x^0$ and $u$. Note that this property \emph{does not depend} on the asymptotical (Schur) stability of the system~\eqref{eq.fjmodel-tv}. For a general neutrally stable system, such a robustness does not hold as illustrated by the simplest counterexample $x(k+1)=x(k)+u$.

Henceforth we are primarily interested in \emph{asymptotically stable} TVFJ models, which means, as usual, that $x(k|x^0,0)\xrightarrow[k\to\infty]{}0$ for any initial condition $x^0$, i.e.
\be\label{eq.product}
\La(k)W(k)\La(k-1)W(k-1)\ldots\La(0)W(0)\xrightarrow[k\to\infty]{} 0.
\ee
Unlike the stationary case, the asymptotical stability in general \emph{does not} imply the convergence~\eqref{eq.x-inf0}.
For instance, let two stationary Schur stable FJ models with matrices $(\La_1,W_1)$ and $(\La_2,W_2)$ corresponding to different matrices $V_1,V_2$ (defined by~\eqref{eq.x-inf}). Due to~\eqref{eq.x-inf}, when $(\La(k),W(k))$ switches between $(\La_1,W_1)$ and $(\La_2,W_2)$ with sufficiently large dwell time, $x(k)$ oscillates between $V_1u$ and $V_2u$. Nevertheless, two ``relaxed'' versions of the convergence condition remain valid for asymptotically stable models~\eqref{eq.fjmodel-tv}.
\begin{lem}\label{lem.stab-tv}
The following conditions are equivalent:
\begin{enumerate}
\item \emph{(stability)} the system~\eqref{eq.fjmodel-tv} is asymptotically stable;
\item \emph{(containment)} for any $x(0),u\in\r^n$ one has
$$
\min_j u_j\le\liminf_{k\to\infty} x_i(k)\le \limsup_{k\to\infty} x_i(k)\le\max_j u_j;
$$
\item \emph{(consensus)} if $u=u_*\ones_n$, then $x(k)\xrightarrow[k\to\infty]{} u\,\forall x(0)$.
\end{enumerate}
\end{lem}

Lemma~\ref{lem.stab-tv} establishes an important relation between the TVFJ model and algorithms of multi-agent
control, namely, protocols for \emph{leader-following} consensus~\citep{RenBeardBook} and
\emph{containment control}~\citep{RenCaoBook}. Adding a ``virtual'' agent $n+1$ whose opinion is static
$x_{n+1}(k)\equiv x_{n+1}(0)$ and the ``augmented'' opinion vector $\hat x(k)=(x_1(k),\ldots,x_n(k),x_{n+1}(k))$,
the system~\eqref{eq.fjmodel-tv} with $u=x_{n+1}(0)\ones_n$ can be rewritten as follows
\be\label{eq.conse}
\hat x(k+1)=\hat A(k)\hat x(k),\; \hat A(k)=\begin{bmatrix}
\La(k) W(k) & (\ones_n-\La(k)\ones_n)\\
0_{1\times n} & 1
\end{bmatrix}.
\ee
Lemma~\ref{lem.stab-tv} states that stability of the model~\eqref{eq.fjmodel-tv} is equivalent
to establishing consensus in~\eqref{eq.conse} $x_i(k)\xrightarrow[k\to\infty]{} x_{n+1}(0)\,\forall i=1,\ldots,n$
for any initial condition $\hat x(0)$ ($i=1,\ldots,n$). This
implies the following stability condition.
\begin{lem}\label{lem.stab-tv-conse}
Suppose that $\ve>0$ exists such that the matrix $\hat A(k)=(\hat a_{ij}(k))$ at any time $k\ge 0$ satisfies the conditions
$\hat a_{ij}(k)\in\{0\}\cup [\ve,1]$ for any $i,j$ and $\hat a_{ii}(k)\ge\ve$ for any $i$.
Then the model~\eqref{eq.fjmodel-tv} is stable if a period $T\ge 1$ exists such that in the graph
$\g[\hat A(k)+\ldots+\hat A(k+T-1)]$ each node is connected to node $n+1$ by a walk.
This holds e.g. if the condition from Lemma~\ref{lem.stab0} is valid at any time.
\end{lem}
\begin{pf}
Thanks to the standard consensus criterion for time-varying directed graphs~\citep{Blondel:05,RenBeardBook},
the assumption of Lemma~\ref{lem.stab-tv-conse} entail consensus in the augmented network~\eqref{eq.conse}, which, in turn, is equivalent to stability of the model~\eqref{eq.fjmodel-tv} due to Lemma~\ref{lem.stab-tv}.
\end{pf}

The assumptions of Lemma~\ref{lem.stab-tv-conse}, typically adopted to prove the convergence of multi-agent coordination
algorithms~\citep{RenBeardBook,RenCaoBook}, are however very restrictive for networks of social influence.
Lemma~\ref{lem.stab-tv-conse}, in particular, is not applicable to TVFJ models where some agents have zero levels of
self-confidence $w_{ii}=0$ or ``totally prejudiced'' $\la_{ii}=0$. Unlike multi-agent \emph{control}
algorithms that are usually designed to have uniformly positive influence weights, such a positivity condition cannot be guaranteed for opinion dynamics. In particular, the process of reflected appraisal~\citep{FriedkinJiaBullo:2016} often leads to the situation where some self-confidence weights $w_{ii}$ asymptotically vanish.

The following two counterexamples demonstrate that in presence of agents with $\la_{ii}(k)w_{ii}(k)=0$
Lemma~\ref{lem.stab-tv-conse} is not valid, in particular, Schur stability of any
matrix $\La(k)W(k)$ does not imply the stability of the model~\eqref{eq.fjmodel-tv}.
We start with two simple counterexamples: in one of them, the matrix $\La$ is fixed while $W$ is switching, in the other one the matrix $W$ is fixed and $\La$ switching.

{\bf Example 1.} Consider $n=3$ agents with $\La(k)\equiv\diag(0,1,1)$ and let the matrix $W(k)$ switch as follows
$$
W(2m)=
\left[\begin{smallmatrix}
1 & 0 & 0\\
1 & 0 & 0\\
0 & 1 & 0
\end{smallmatrix}\right],\, W(2m+1)=
\left[\begin{smallmatrix}
1 & 0 & 0\\
0 & 0 & 1\\
1 & 0 & 0
\end{smallmatrix}\right],\quad m=0,1,\ldots.
$$
The dynamics~\eqref{eq.fjmodel-tv} implies that $x_1(k)\equiv u_1$ and
\ben
(x_2(k+1),x_3(k+1))=\begin{cases}
(u_1,x_2(k)),\quad k=2m,\\
(x_3(k),u_1),\quad k=2m+1.
\end{cases}
\een
Therefore, we have $x_2(0)=x_2(2)=\ldots=x_2(2m)\,\forall m$ and $x_2(k)\not\to 0$ as $k\to\infty$ when $u_1=0$ and $x_2(0)\ne 0$.

{\bf Example 2.} Consider the TVFJ model with $n=2$ and
\ben
W(k)\equiv \left[\begin{smallmatrix}
0 & 1\\
1 & 0
\end{smallmatrix}\right],\,\La(2m)=
\left[\begin{smallmatrix}
0 & 0\\
0 & 1
\end{smallmatrix}\right],
\La(2m+1)=
\left[\begin{smallmatrix}
1 & 0\\
0 & 0
\end{smallmatrix}\right].
\een
The dynamics~\eqref{eq.fjmodel-tv} can then be rewritten as follows
\ben
(x_1(k+1),x_2(k+1))=\begin{cases}
(u_1,x_2(k)),\quad k=2m,\\
(x_1(k),u_2),\quad k=2m+1,
\end{cases}
\een
entailing that $x_1(0)=x_1(2)=\ldots=x_1(2m)\,\forall m$,
and thus $x_1(k)\not\to 0$ as $k\to\infty$ when $u=0$ and $x_1(0)\ne 0$.

In Examples~1 and 2 the switching model~\eqref{eq.fjmodel-tv} appears to be not asymptotically stable in spite of the Schur stability of the two possible values $\La(k)W(k)$: the
joint spectral radius~\citep{LinAntsaklis2009} of these matrices equals to $1$. This critical situation, where the results of classical switching systems theory~\citep{LinAntsaklis2009} are not applicable, is typical for the TVFJ model. To guarantee its stability, special criteria are needed; one of such criteria, extending Theorem~\ref{thm.stab}, is offered in this section.

We start with introducing a class $\cfj[\delta,\ve,s]$, where $\delta,\ve>0$ are real and $s\ge 0$
is an integer (acronym CFJ stands for ``Chain of FJ models''). Unlike $\fj[\delta,\ve,s]$,
constituted by pairs $(\La,W)$, the class $\cfj[\delta,\ve,s]$ consists of sequences $\{(\La^{(k)},W^{(k)})\}_{k=0}^{s}$.
For such a sequence and $\delta,\ve>0$, we introduce the sets
\ben
J_0\dfb S_{\La^{(0)}}^{\delta},\; J_{k}\dfb S_{\La^{(k)}}^{\delta}\cup \{i:w^{(k)}_{ij}\ge\ve \text{ for some $j\in J_{k-1}$}\}.
\een
When $\La^{(k)}=\La$ and $W^{(k)}=W$ for any $k=0,\ldots,s$, the set $J_j$ contains all such indices $i$ that $d_{\ve}(i,S_{\La}^{\delta},W)\le j$.
\begin{defn}
The class $\cfj[\delta,\ve,s]$ consists of all sequences $\{(\La^{(k)},W^{(k)})\}_{k=0}^{s}$ such that $J_s=\{1,2,\ldots,n\}$.
\end{defn}

The following result is proved similarly to Theorem~\ref{thm.stab}.
\begin{lem}\label{lem.chain}
For any sequence $\{(\La^{(k)},W^{(k)})\}_{k=0}^{s}$ from the set $\cfj[\delta,\ve,s]$ the matrix $\mathcal{P}\dfb\prod_{k=0}^s\La^{(k)}W^{(k)}$ has row sums $\le 1-\delta\ve^s$, that is,
$\mathcal{P}\ones_n\le (1-\delta\ve^s)\ones_n$.
\end{lem}

Using Lemma~\ref{lem.chain}, the following sufficient condition for asymptotic stability is immediate.
\begin{thm}\label{thm.stab-tv}
Let real $\delta,\ve>0$ and an integer $s\ge 0$ exist such that the sequence $\{(\La(k),W(k))\}_{k=0}^{\infty}$ contains infinitely many subsequences $\{(\La(k),W(k))\}_{k=m}^{m+s}$ from $\cfj[\delta,\ve,s]$. Then the model~\eqref{eq.fjmodel-tv} is asymptotically stable.
\end{thm}

The condition of Theorem~\ref{thm.stab-tv} can, evidently, be reformulated as follows: any infinite ``tail'' $\{(\La(k),W(k))\}_{k=r}^{\infty}$ (where $r\ge 1$) contains a subsequence from $\cfj[\delta,\ve,s]$. This condition does not require stability of any matrix $\La(k)W(k)$ and allows e.g. to have $\La(k)=I_n$ for some $k$.

\section{Proofs}
In this section, we prove our main results.

\subsection{Proof of Theorem~\ref{thm.cons}}
We start with the sufficiency part. Suppose that $u_i=u_0\,\forall i\in S_{\La}$. One may assume that $u_i=u_0\,\forall i$ since for $i\not\in S_{\La}$ the value of $u_i$
has no effect on $x^{\infty}$. Since $V$ is row-stochastic~\citep{Friedkin:2015}, $x^{\infty}=V(u_0\ones_n)=u_0\ones_n$, which proves consensus. To prove necessity, assume that $x^{\infty}=u_0\ones_n$ for some $u_0\in\r$. Using~\eqref{eq.x-inf}, $(I_n-\La)u=u_0(I_n-\La W)\ones_n=u_0(I_n-\La)\ones_n$. In view of~\eqref{eq.s}, $u_i=u_0\,\forall i\in S_{\La}$. $\blacksquare$

\subsection{Proofs of Theorems~\ref{thm.stab},\ref{thm.stab-tv}, Lemma~\ref{lem.chain} and Corollary~\ref{cor.stab}}

We start with a useful technical lemma. Given a substochastic matrix $A$, the number $\hat a_i=1-\sum_{j}a_{ij}\ge 0$ is said to be the \emph{deficiency} of the $i$th row.
From the Gershgorin disk theorem~\citep{HornJohnsonBook1985} it is obvious that $\rho(A)\le\max\limits_{1\le i\le n}[1-\hat a_i]$.
\begin{lem}
Let $A$ and $B$ be substochastic $n\times n$ matrices, $C=AB$ and $\hat a_i,\hat b_i,\hat c_i$ stand for the respective deficiencies. Then the following statements hold for any $i=1,\ldots,n$
\begin{gather}
\hat c_i=\hat a_i+\sum_{j=1}^na_{ij}\hat b_j\label{eq.lem-1}.
\end{gather}
\end{lem}
\begin{pf}
Denote the $i$th row of $B$ and $C$ with respectively $b_{i\bullet}$ and $c_{i\bullet}$, we have
$c_{i\bullet}\ones_n=\sum_{j}a_{ij}b_{i\bullet}\ones_n=\sum_ja_{ij}(1-\hat b_j)=1-\hat a_i-\sum_ja_{ij}\hat b_j$, which entails~\eqref{eq.lem-1}.
\end{pf}

\emph{Proof of Theorem~\ref{thm.stab}.} For brevity, we denote $A\dfb\La W$ and put $d(i)\dfb d_{\ve}(i,S_{\La}^{\delta},W)$.
We are going to prove the following statement via induction on $m$: if $d(i)\le m$ and $C=A^{1+m}$, then $\hat c_i\ge\delta\ve^{d(i)}$.
For $m=0$ the claim is obvious: when $d(i)=0$ one has $i\in S_{\La}^{\delta}$ and hence $\hat c_i=\hat a_i\ge\delta$.
Assuming that the statement has been proved for $m-1\ge 0$, we have to prove it for $m$. Denoting $B=A^m$, one has
$C=BA$. If $d(i)\le m-1$ then $\hat c_i\ge\hat b_i\ge\delta\ve^{d(i)}$ thanks to~\eqref{eq.lem-1}. If $d(i)=m$, there exists $j$ such that $d(j)\le m-1$ and $w_{ij}\ge\ve$.
Denoting $C'=WB$, \eqref{eq.lem-1} implies that $\hat c_i'\ge\ve\hat b_i\ge\delta\ve^m$.
Since $C=\La C'$, we have $\hat c_i\ge\hat c_i'$ which proves our statement for $m$.
Substitution $m=s$ yields $\rho(A^{1+s})=\rho(A)^{1+s}\le 1-\delta\ve^s$ by definition of $\fj(\delta,\ve,s)$.$\blacksquare$

\emph{Corollary~\ref{cor.stab}} is immediate from Theorem~\ref{thm.stab} since a stable FJ model belongs to $\fj[\delta_0,\ve_0,n-1]$.$\blacksquare$

\emph{Proof of Lemma~\ref{lem.chain}.} Similarly to proof of Theorem~\ref{thm.stab}, one proves via induction on $m=0,1,\ldots,s$ that for any $i\in J_m$ the $i$th row of the matrix
$
\mathcal P_{m}\dfb\prod_{k=0}^m\La^{(k)}W^{(k)}
$ has deficiency $\ge\delta(1-\delta)^m\ve^m$. For $m=0$ the claim is trivial. Assuming that it holds for $m-1$, let $A=\La^{(m)}W^{(m)}$ and $B=\mathcal P_{m-1}$.
If $i\in J_m$ then either $i\in S_{\La^{(m)}}^\delta$ and hence $\hat a_i\ge\delta$ or such $j$ exists for which $w_{ij}^{(m)}\ge \ve$ and $\hat b_j\ge \delta\ve^{m-1}$.
Using~\eqref{eq.lem-1}, one now proves the claim for $m$ in the same way as in Theorem~\ref{thm.stab}.$\blacksquare$

\emph{Proof of Theorem~\ref{thm.stab-tv}} is immediate from Lemma~\ref{lem.chain}. Let $P(m)\dfb\prod_{k=0}^m\La(k)W(k)$. Notice that if $P(m-1)\ones_n\le \theta\ones_n$ and $\{(\La(k),W(k))\}_{k=m}^{m+s}$ belongs to $\cfj[\delta,\ve,s]$, then $P(m+s)\ones_n\le \theta(1-\delta\ve^s)\ones_n$ due to Lemma~\ref{lem.chain}. This implies, via induction on $r=0,1,\ldots$, that if the sequence $\{(\La(k),W(k))\}_{k=0}^{m}$ contains $r$ non-intersecting subsequences from $\cfj[\delta,\ve,s]$, then one has $P(m)\ones_n\le (1-\delta\ve^s)^r$. Therefore $P(m)\to 0$ as $m\to\infty$.$\blacksquare$

\subsection{Proofs of Lemmas~\ref{lem.simple} and \ref{lem.stab-tv}}

\emph{Proof of Lemma~\ref{lem.simple}.} Statements~1)-3) are proved using induction on $k$. For instance, if $x_i(0)\le M$ and $u_i\le M$ for any $i$, then $x_i(1)\le M\,\forall i$ due to~\eqref{eq.fjmodel-tv}. Therefore, $x_i(2)\le M$ and so on. Statement~4) follows from 2) due to the linearity of~\eqref{eq.fjmodel-tv}: $x(\cdot|x^0+\delta x,u+\delta u)=x(\cdot|x^0,u)+x(\cdot|\delta x,\delta u)$.$\blacksquare$

\emph{Proof of Lemma~\ref{lem.stab-tv}.} Implications 2)$\Longrightarrow$3) and 3)$\Longrightarrow$1) are obvious (the first of them is proved by putting $u_i=u_*$ and the second one by taking $u_*=0$). To prove the implication 1)$\Longrightarrow$2), note that the limits $\limsup x(k)$, $\liminf x(k)$ do not depend on $x(0)$ due to stability.
Assuming that $x(0)=0$, the claim follows now from statement 2) in Lemma~\ref{lem.simple} by substituting $m=\min_iu_i$ and $M=\max_iu_i$.$\blacksquare$

\section{Conclusions}

In this paper, important system-theoretic properties of the Friedkin-Johnsen (FJ) model of opinion dynamics~\citep{FriedkinJohnsen:1999} are considered such as stability and convergence speed. We also examine the extension of the FJ model to the case of time-varying social influence and give sufficient conditions for its stability.
The time-varying FJ model can be further extended to the case of multidimensional opinions, representing the agents'
positions on several interrelated issues (\emph{belief systems}); for static FJ model such an extension is discussed
in~\citep{Parsegov2017TAC,FriedkinPro2016}.
In our future works we are going to validate the applicability of the FJ model to opinion dynamics in large-scale online social networks.

\bibliography{consensus}

\end{document}